\documentclass{mem}
\usepackage{natbib}\usepackage{txfonts}\usepackage{balance}
\usepackage{graphicx}
\usepackage[a4paper,breaklinks,dvipdfm]{hyperref}
\idline{1}{1}
\begin{document}
\def\teff{$T\rm_{eff }$}
\def\kms{$\mathrm {km s}^{-1}$}

\title{
X-ray binaries powered by massive stellar black holes
}

   \subtitle{}

\author{
Michela \,Mapelli
          }


\institute{
INAF --
Osservatorio Astronomico di Padova, Vicolo dell'Osservatorio 5,
I--35122 Padova, Italy, \email{michela.mapelli@oapd.inaf.it}
}

\authorrunning{Mapelli}

\titlerunning{X-ray binaries powered by MSBHs}

\abstract{
The mass of stellar black holes (BHs) is currently thought to be in the $3-20$ M$_\odot$ range, but  is highly uncertain: recent observations indicate the existence of at least one BH with mass $>20$ M$_\odot$.  The metallicity of the progenitor star strongly influences the mass of the remnant, as only metal-poor stars can have a final mass higher than $\sim{}40$ M$_\odot$, and are expected to directly collapse into BHs with mass $>25$ M$_\odot$.  By means of N$-$body simulations, we investigate the formation and evolution of massive stellar BHs (MSBHs, with mass $>25$ M$_\odot$) in young dense star clusters.  We study the effects of MSBHs on the population of X-ray sources. 
}
\maketitle{}

\section{Introduction}
\begin{figure}[]
\resizebox{\hsize}{!}{\includegraphics[clip=true]{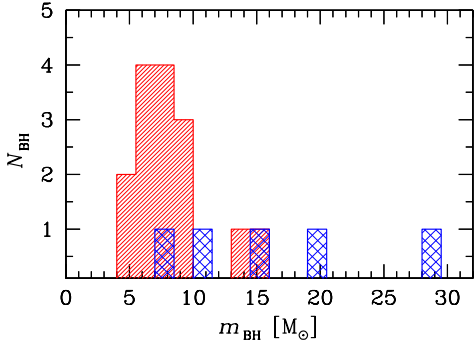}}
\caption{
\footnotesize
Distribution of BH masses derived from dynamical measurements. The BH masses were taken from the literature and are listed in Table~1. 
Hatched red histogram: Milky Way BHs; cross-hatched blue histogram: BHs in nearby galaxies.} 
\label{fig:fig1}
\end{figure}

The mass spectrum of stellar black holes (BHs) is currently thought to be in the $3-20$ M$_\odot$ range, but is highly uncertain. Dynamical mass estimates are available only for a few tens of BHs, hosted in X-ray binaries (see table 2 of \citealt{ozel10} for one of the most updated compilations). In the Milky Way, the dynamically measured BH masses range from $\approx{}4$ to  $\approx{}15$ M$_\odot{}$. 
BHs with mass $>15$ M$_\odot{}$ were found in some nearby galaxies.
In Table~1 and in Fig.~\ref{fig:fig1}, we report a compilation of BH masses from the literature. All of them come from dynamical measurements, and are among the best constrained values. Three out of five BHs in nearby galaxies (last 5 lines of Table~1) have mass $\gtrsim{}15$ M$_\odot{}$. In the case of IC~10 X-1, the BH mass can be as high as $\approx{}30$ M$_\odot{}$ (e.g., \citealt{prestwich07}). 

Which factors can affect the BH mass? Theoretical models (e.g. \citealt{heger03}; \citealt{mapelli09}; \citealt{zampieri09}; \citealt{belczynski10}; \citealt{fryer12}) indicate that the metallicity ($Z$) of the progenitor star can significantly influence the mass of the BH. In particular, a massive star can collapse quietly into a BH (i.e. without supernova or with a faint supernova), if its final mass is sufficiently high ($\approx{}40$ M$_\odot$, \citealt{fryer99}). Massive metal-poor stars lose less mass by stellar winds than metal-rich stars (e.g. \citealt{vink01}), and thus are more likely to have a final mass $\gtrsim{}40$ M$_\odot$. The mass of a BH born from direct collapse is expected to be very close to the final mass of the progenitor star: it can significantly exceed $25$ M$_\odot{}$, depending on the metallicity. 
Interestingly, the stellar BH with the highest dynamically measured mass
is hosted in the metal-poor galaxy IC~10 ($Z = 0.22$ Z$_\odot{}$, from an electron-temperature based calibration of spectra of HII regions, assuming Z$_\odot{}$ = 0.019).  

In the following, we call massive stellar BHs (MSBHs) those BHs with mass $>25$
 M$_\odot{}$, born from a quiet collapse. The existence of MSBHs in the nearby
Universe may be crucial for our understanding of X-ray
sources. The scenario of X-ray binaries powered by MSBHs
was recently proposed to explain a fraction of the ultraluminous
X-ray sources (ULXs, i.e. point-like X-ray sources with luminosity,
assumed isotropic, higher than $10^{39}$ erg s$^{-1}$, 
e.g. \citealt{mapelli10}).
In this paper, we study (by means of N$-$body simulations) the formation and evolution of MSBHs in young star clusters (SCs), and we investigate their importance for the population of accreting binaries.

\begin{table}
\caption{Compilation of BH masses from dynamical measurement.}
\begin{tabular}{lccc}
\hline
Name             & Type$^{\rm a}$ & $m_{\rm BH}$ (M$_\odot{}$) & Reference$^{\rm b}$ \vspace{0.3cm}\\
\hline
GRO J 0422+324.3 & SPT & $3.7-5.0$      & (1) \\
A 0620-003       & SPT & $6.6\pm{}0.2$  & (2) \\
GRS 1009-45      & SPT & $3.6-4.7$      & (1) \\
XTE J 1118+480   & SPT & $6.5-7.2$      & (1) \\
GS 1124-683      & SPT & $6.5-8.2$      & (1) \\
GS 1354-64       & LPT & $7.9\pm{}0.5$  & (3) \\ 
4U 1543-47       & LPT & $9.4\pm{}1.0$  & (2) \\
XTE J 1550-564   & LPT & $9.1\pm{}0.6$  & (2) \\
GRO J 1655-40    & LPT & $6.3\pm{}0.3$  & (2) \\
H 1705-250       & SPT & $5.6-8.3$      & (1) \\  
SAX J 1819.3-2525 & LPT & $7.1\pm{}0.3$  & (2) \\
GRS 1915+105     & LPT & $14\pm{}4$     & (4) \\
Cyg X-1          & PS  & $14.8\pm{}1.0$ & (5) \\
GS  2000+251     & SPT & $7.1-7.8$      & (1) \\
GS  2023+338     & LPT & $9.0^{+0.2}_{-0.6}$     & (6) \vspace{0.4cm} \\
IC 10  X-1       & PS  & $24-33$        & (7) \\
NGC300 X-1       & PS  & $20\pm{}4$     & (8) \\
M33 X-7          & PS  & $15.65\pm{}1.45$ & (9) \\
LMC X-3          & PS  & $5.9-9.2$      & (1) \\
LMC X-1          & PS  & $10.91\pm{}1.54$ & (10) \\ 
\hline
\end{tabular}
\caption{
\footnotesize
$^{\rm a}$SPT (LPT): short- (long-) period transient; PS: persistent source. See \citet{ozel10}.
$^{\rm b}$References: (1) \citet{orosz03}, and references therein; (2) \citet{ozel10}, and references therein; (3) \citet{casares09}; (4) \citet{harlaftis04}; (5) \citet{orosz11}; (6) \citet{khargharia10}; (7) \citet{prestwich07}; (8) \citet{crowther10}; (9) \citet{orosz07}; (10) \citet{orosz09}.}
\end{table}



\section{Simulations}
We simulate young SCs, as (i) most stars form in SCs (\citealt{lada03}), and (ii) high-mass X-ray binaries and ULXs are often associated with OB associations and with young SCs (e.g.  \citealt{zezas02}; \citealt{soria05}; 
\citealt{swartz09}). 

Most SCs are collisional environments: their two-body relaxation timescale is short with respect to their lifetime. 
Thus, binaries in SCs undergo a number of three-body encounters, i.e. close encounters with single stars.
Three-body encounters affect the population of accreting binaries, in collisional environments. For example, three-body encounters can change the semi-major axis of a binary, triggering mass transfer. Furthermore, a BH that was born from a single star can become member of a binary, by replacing one of its former members through a dynamical exchange.

We perform N-body simulations of SCs using the Starlab public software environment (\citealt{portegies01}). We modified Starlab, to include metal-dependent stellar evolution and recipes for stellar winds by \citeauthor{vink01} (\citeyear{vink01}; see also \citealt{mapelli13} for more details on the code).
We simulated young intermediate-mass SCs, generated according to a multi-mass King model, with total mass $M_{\rm TOT}=3000-4000$ M$_\odot{}$, initial core radius $r_{\rm c}=0.4$ pc, concentration $c=1.03$. The stars in the SC follow a \citet{kroupa01} initial mass function. We include a fraction $f_{\rm b}=0.1$ of primordial binaries in the initial conditions. We consider three different metallicities: $Z=0.01$, 0.1 and 1 Z$_\odot{}$. We ran 100 realizations of the same SC (by changing the random seed) for each $Z$, to filter out the statistical fluctuations.


\section{Results}
\begin{figure}[]
\resizebox{\hsize}{!}{\includegraphics[clip=true]{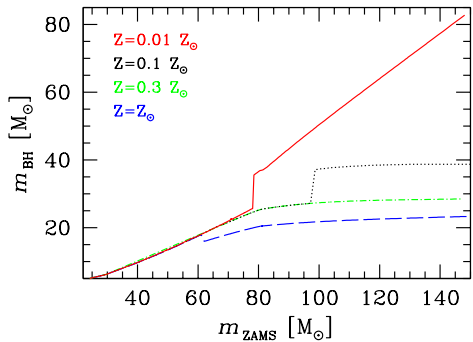}}
\caption{
\footnotesize
Mass of the BH ($m_{\rm BH}$) versus ZAMS mass ($m_{\rm ZAMS}$) of the progenitor star, for a population of single stars. 
Solid red line: 0.01 Z$_\odot$; dotted black line:
0.1 Z$_\odot$; dot-dashed green line: 0.3 Z$_\odot$; dashed blue line: 1 Z$_\odot$.}
\label{fig:fig2}
\end{figure}

Fig.~\ref{fig:fig2} shows the mass spectrum of BHs in our simulations, as a function of the zero-age main sequence (ZAMS) mass of the progenitor star.  The effect of metallicity is apparent: no BHs with mass $>25$ M$_\odot{}$ form at solar metallicity from the evolution of single stars, while MSBHs with mass as high as $\sim{}80$ M$_\odot$ can form at $Z=0.01$ Z$_\odot{}$.

\begin{figure*}[]
\includegraphics[width=4.3cm]{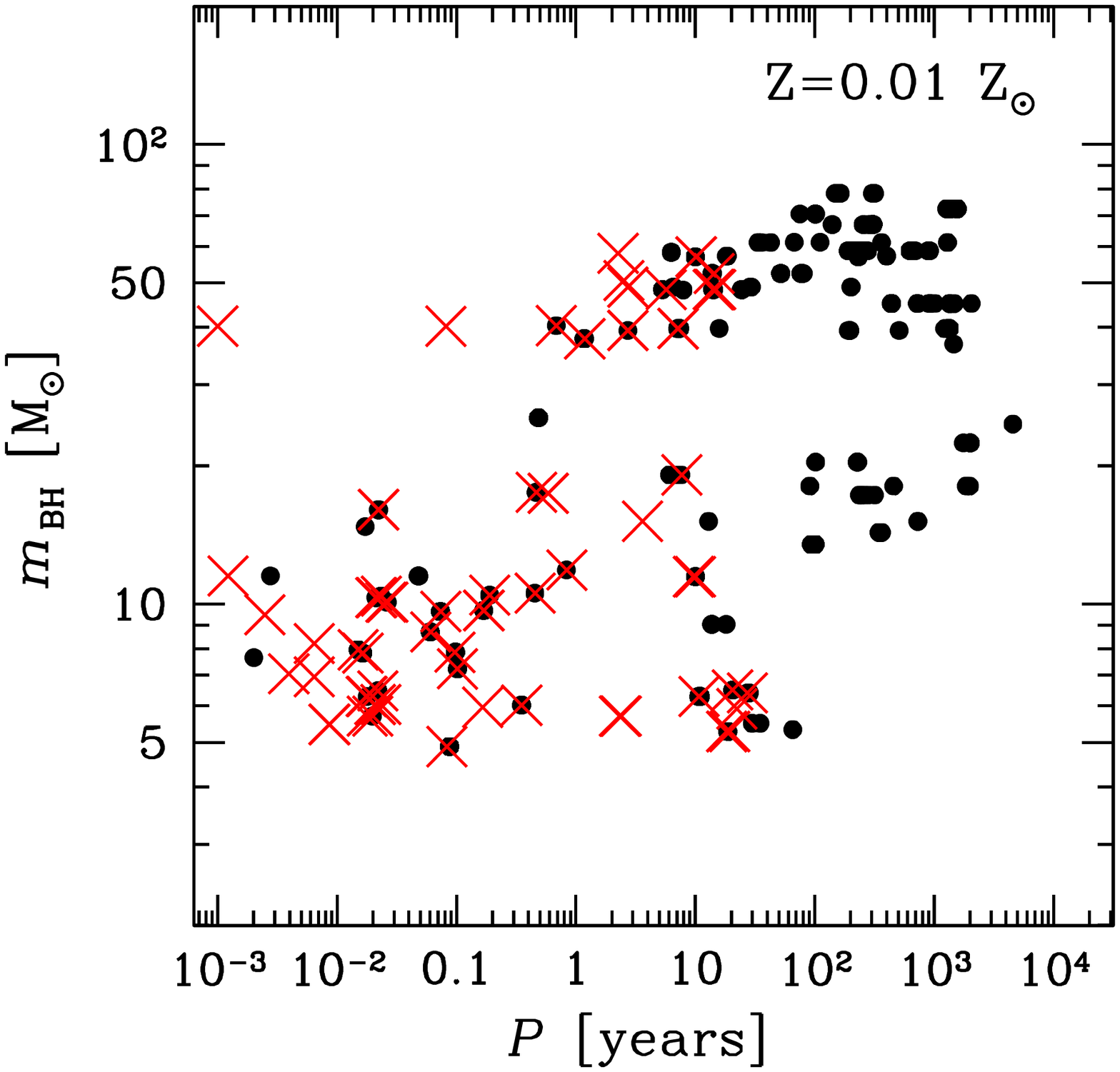}
\includegraphics[width=4.3cm]{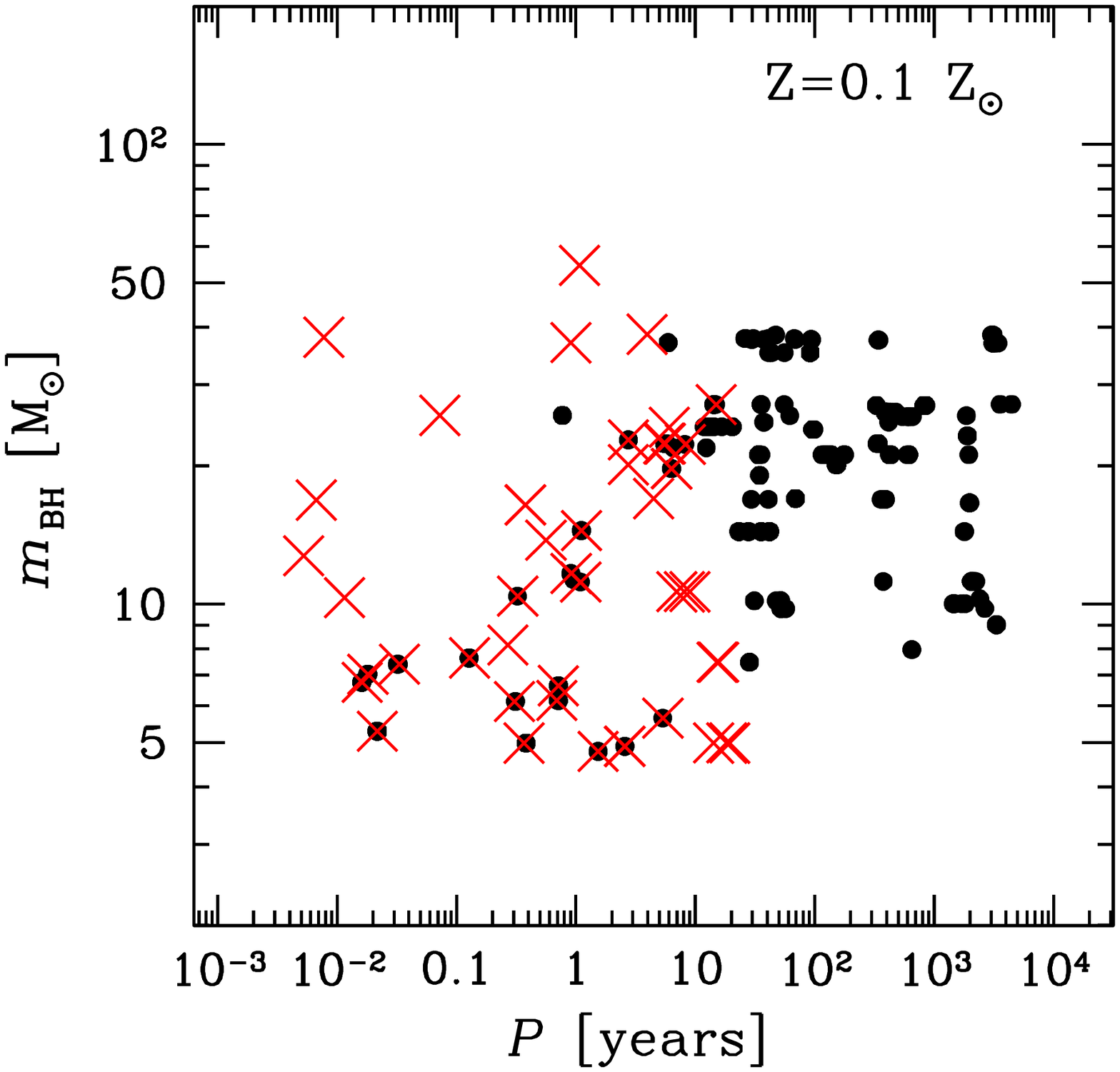}
\includegraphics[width=4.3cm]{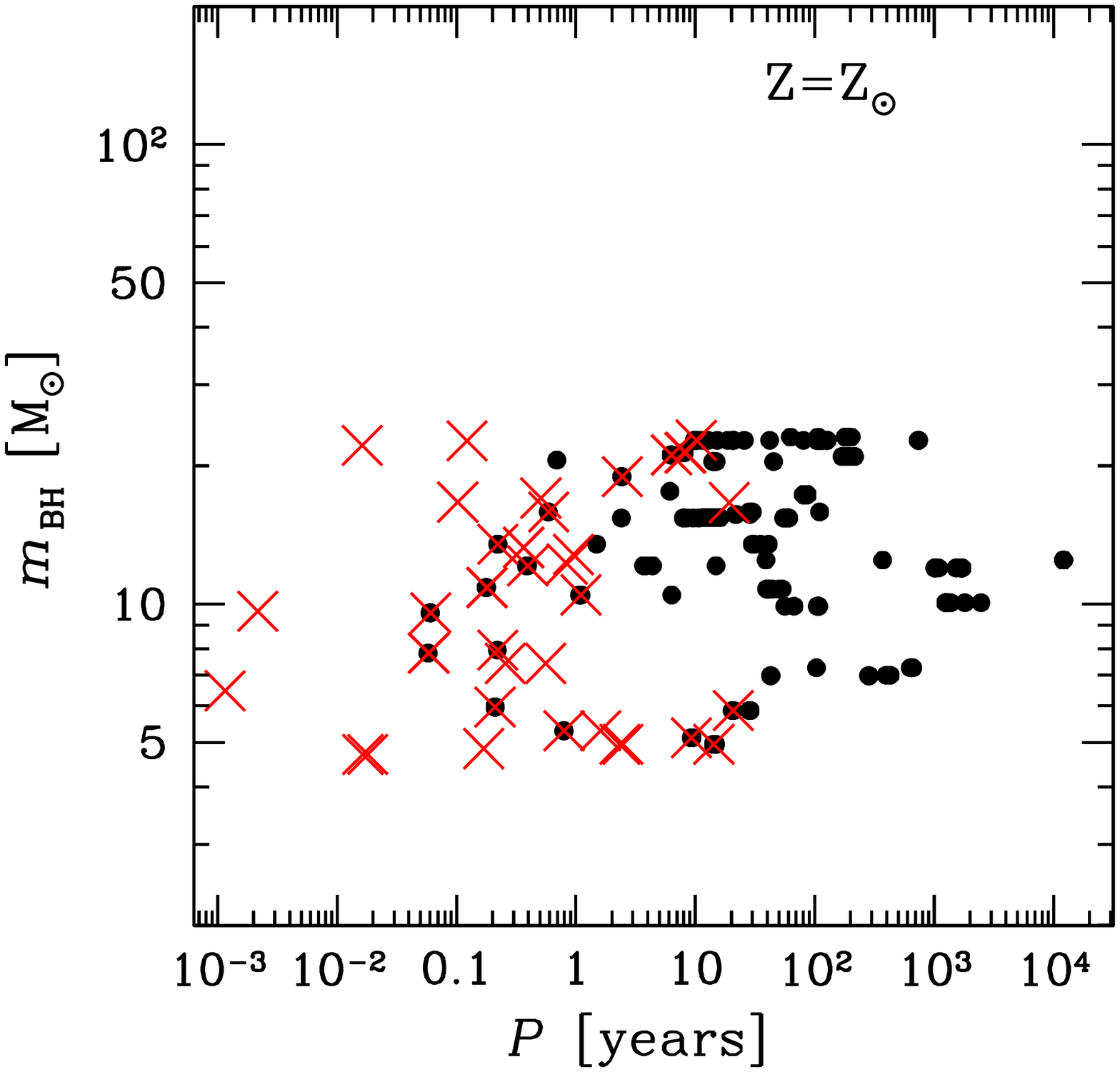}
\caption{\footnotesize Mass of the BH versus orbital period, for the simulated BH binaries, during the accretion phase. Filled circles: wind-accretion systems; red crosses: RLO systems.
 From left to right:  0.01 Z$_\odot{}$,  0.1 Z$_\odot{}$, 1  Z$_\odot{}$. 
}
\label{fig:fig3}
\end{figure*}

\begin{figure*}[]
\includegraphics[width=4.3cm]{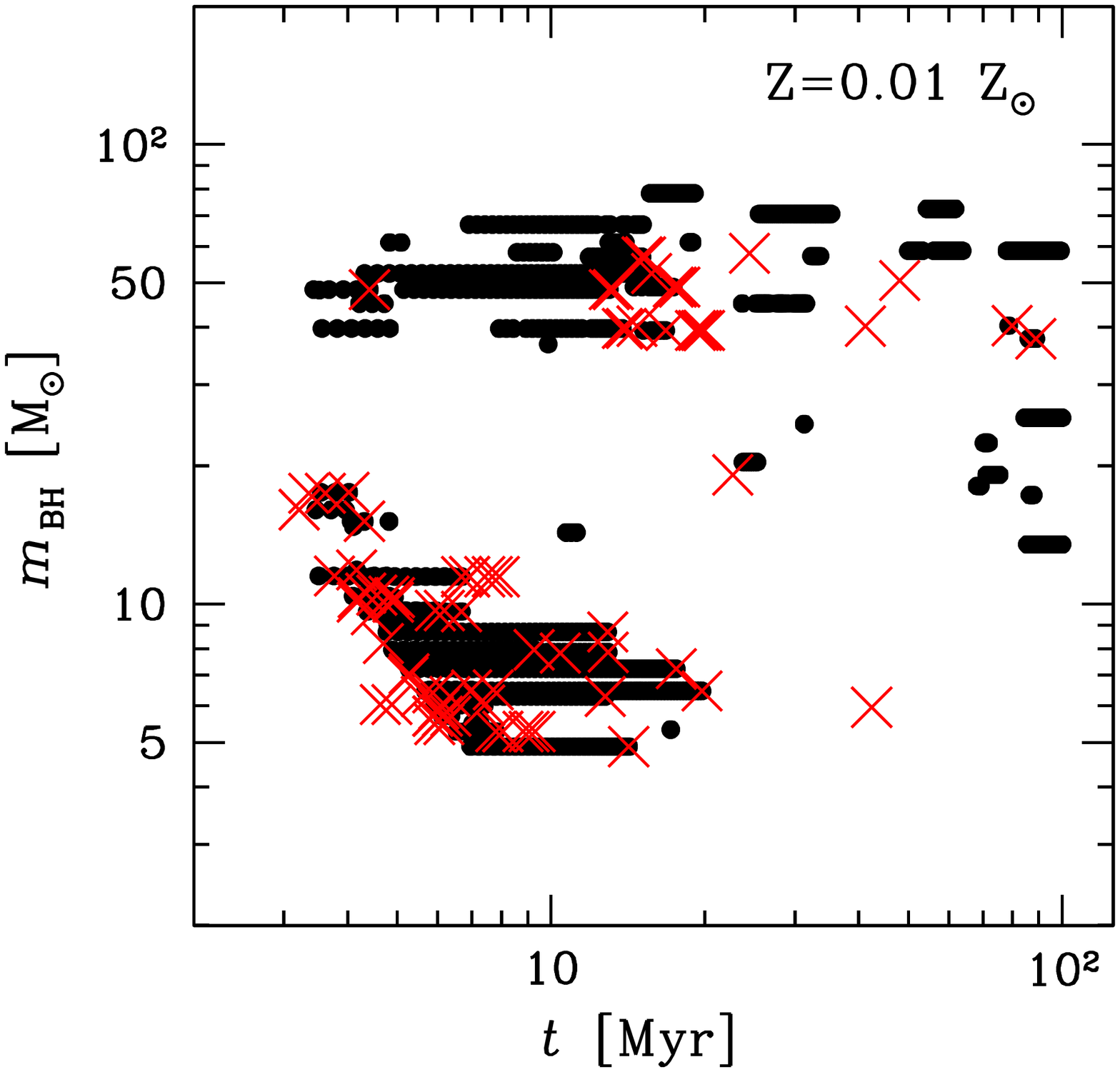}
\includegraphics[width=4.3cm]{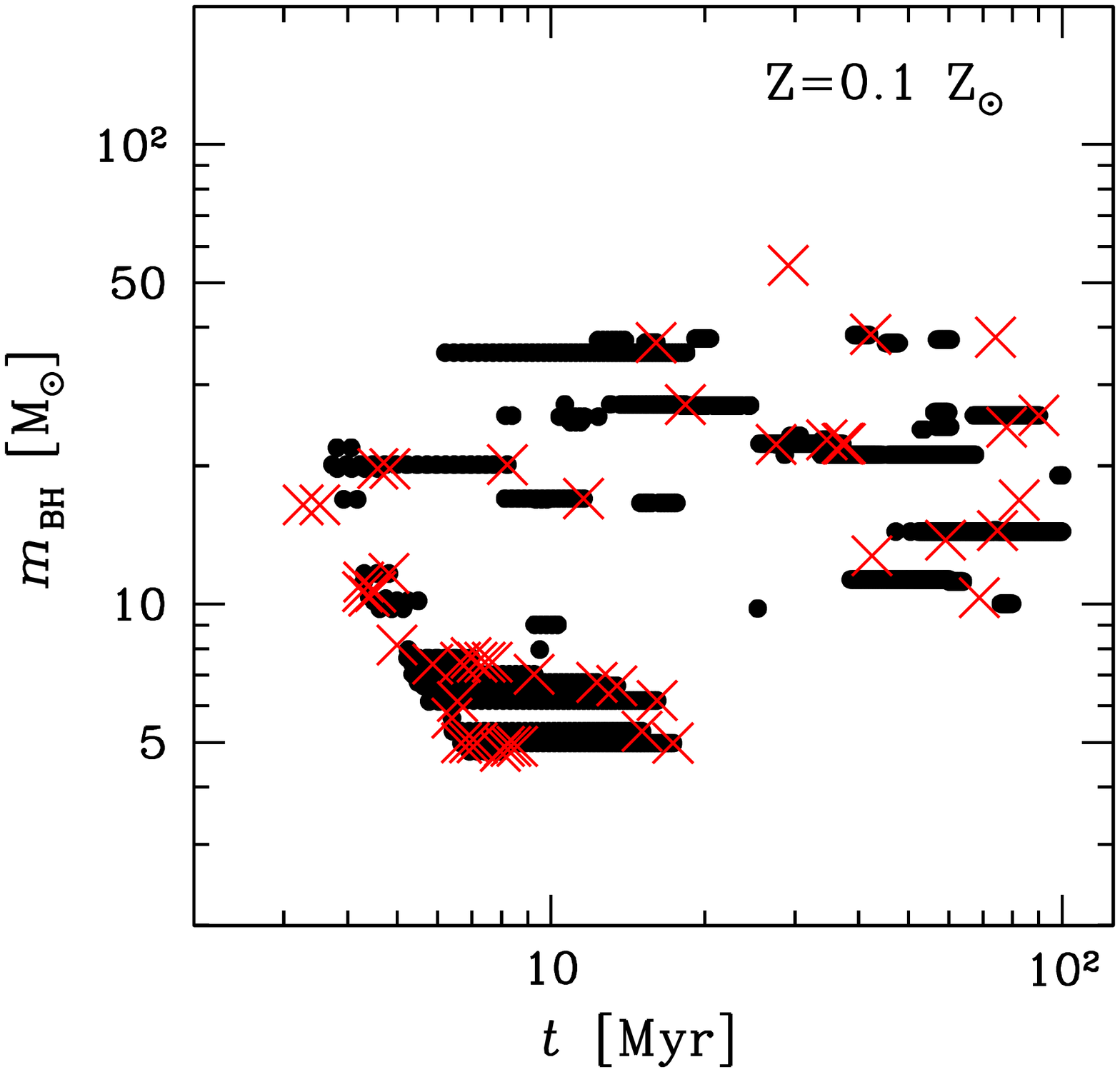}
\includegraphics[width=4.3cm]{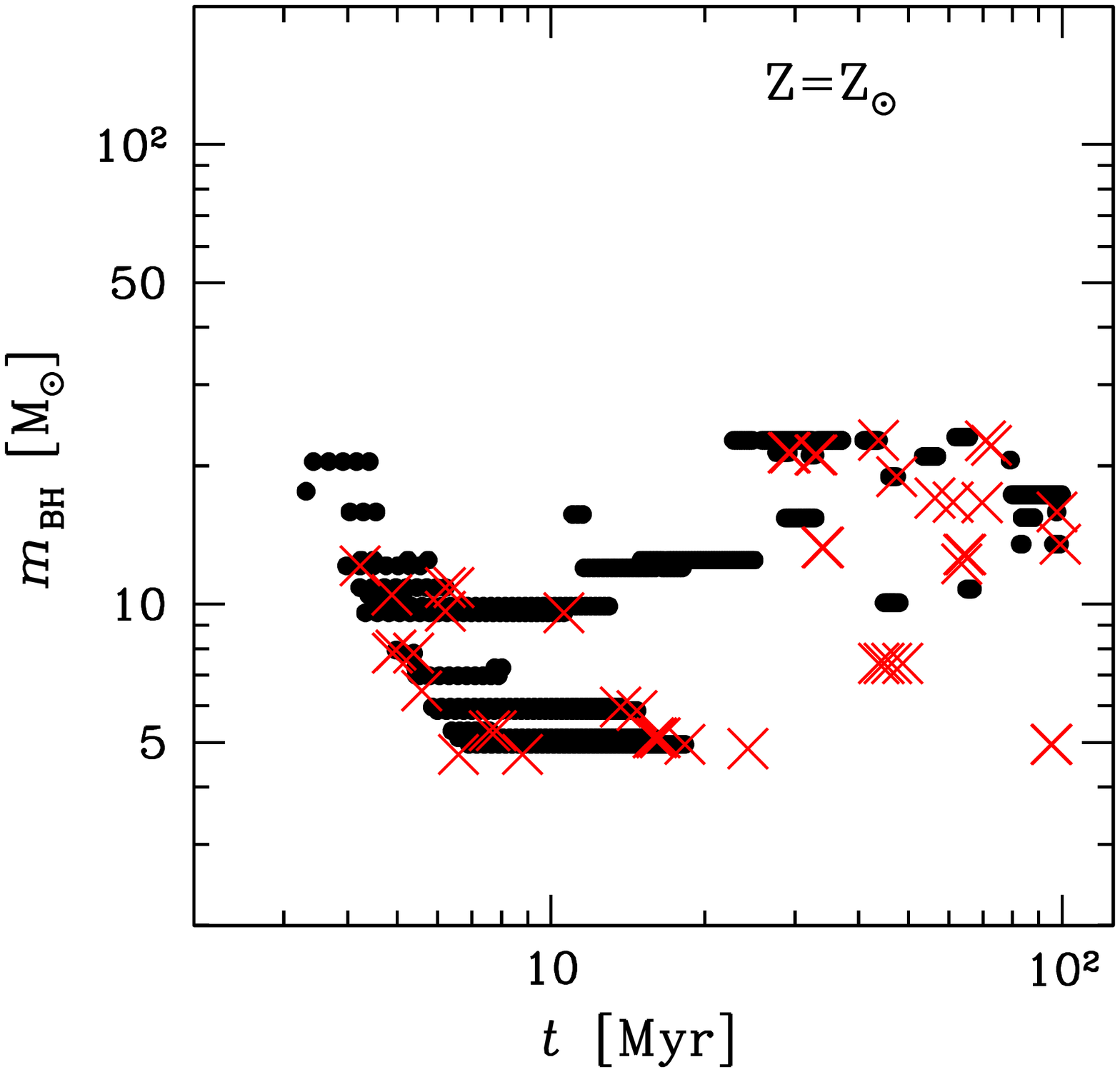}
\caption{\footnotesize
 Mass of the BH versus time elapsed since the beginning of the simulation, for the BH binaries, during the accretion phase. The symbols are the same as in Fig.~\ref{fig:fig3}. 
}
\label{fig:fig4}
\end{figure*}

The simulations provide information about the formation and evolution of X-ray binaries, powered by both Roche lobe overflow (RLO) and wind accretion. Fig.~\ref{fig:fig3} shows the mass of the BH versus the orbital period of RLO (crosses) and wind-accreting (circles) binaries. It is apparent that accreting binaries at low metallicity can be powered by MSBHs: $10-20$ per cent ($5-10$ per cent) of all MSBHs power wind-accreting (RLO) systems. Interestingly, the vast majority ($>90$ per cent) of accreting binaries powered by MSBHs underwent at least one dynamical exchange before starting the accretion. In most cases, the MSBH formed from a single star and then became member of a binary via exchange. This is in agreement with the fact that the rate of three-body encounters (and especially of exchanges) scales with the mass of the involved objects.

 Fig.~\ref{fig:fig4} gives information about when a  binary starts wind accretion or RLO, with respect to the time elapsed since the beginning of the simulation. Binary systems powered by low-mass BHs tend to start the RLO phase earlier  than those powered by MSBHs. The reason is that the RLO phase in systems powered by low-mass BHs is driven by the stellar evolution of the  companion, while in  systems powered by MSBHs it is mainly a consequence of dynamical exchanges, which occur on a longer timescale.

\section{Conclusions}
In this paper, we investigated the importance of MSBHs for the population of accreting binaries in young SCs. We showed that $\sim{}5-10$ per cent of all the simulated MSBHs  power RLO systems. The vast majority of accreting binaries powered by MSBHs underwent at least one dynamical exchange before starting the accretion. Instead, the number of accreting MSBHs in unperturbed primordial binaries is negligible. The key result of our simulations is that MSBHs are efficient in powering X-ray binaries through dynamical evolution. This result indicates that MSBHs can power X-ray binaries in low-metallicity young SCs, and is very promising to explain the association of many ultraluminous X-ray sources with low-metallicity and 
star forming environments. 
\begin{acknowledgements}
 We thank the developers of Starlab, and especially  P. Hut, S. McMillan, J. Makino, and S. Portegies Zwart.  
We acknowledge the CINECA Award N.  HP10CXB7O8 and HP10C894X7, 2011. MM acknowledges financial support from INAF through grant PRIN-2011-1. 
\end{acknowledgements}

\bibliographystyle{aa}

\end{document}